# Holistic framework to help students learn effectively from research-validated self-paced learning tools

Emily Marshman[1], Seth DeVore[2], and Chandralekha Singh[2]
[1]*Department of Physics, Community College of Allegheny County, Pittsburgh, PA 15212*
[2]*Department of Physics and Astronomy, University of Pittsburgh, Pittsburgh, PA 15260*

**Abstract:** With limited time available in the classroom, e-learning tools can supplement in-class learning by providing opportunities for students to study and learn outside of class. Such tools can be especially helpful for students who lack adequate prior preparation. However, one critical issue is ensuring that students, especially those in need of additional help, engage with the tools as intended. Here we first discuss an empirical investigation in which students in a large algebra-based physics course were given opportunities to work through research-validated tutorials outside of class as self-study tools. Students were provided these optional tutorials after traditional instruction in relevant topics and were then given quizzes that included problems that were identical to the tutorial problems with regard to the physics principles involved but had different contexts. We find that students who worked through the tutorials as self-study tools struggled to transfer their learning to solve problems that used the same physics principles. On the other hand, students who worked on the tutorials in supervised, one-on-one situations performed significantly better than them. These empirical findings suggest that many introductory physics students may not engage effectively with self-paced learning tools unless they are provided additional incentives and support, e.g., to aid with self-regulation. Inspired by the empirical findings, we propose a holistic theoretical framework to help create learning environments in which students with diverse backgrounds are provided support to engage effectively with self-study tools.

## I. INTRODUCTION

With limited time available in the classroom, self-paced learning tools provide a valuable opportunity to supplement learning even in brick and mortar classrooms [1-9]. Adaptive learning tools allow students to obtain feedback and support based upon their needs, and students can work through them at their own pace and receive additional help as needed [10-16]. Appropriate use of these learning tools can be particularly beneficial for students with inadequate prior preparation and provide all students an opportunity to learn. These tools can play a central role in scaffolding student learning and helping them learn content as well as develop their problem-solving, reasoning and meta-cognitive skills [17-52].

However, ensuring that students engage effectively with these learning tools is challenging, especially among students who are struggling with the course material and need out of class help. For example, students may not feel confident or lack the self-regulation and time-management skills necessary for effective engagement with these tools [38-48]; moreover, students' environmental constraints may not be conducive to engagement with these tools without additional support. Thus, without sufficient help and incentive to ensure effective engagement with the tools, students may not follow the guidelines for effectively using the tools regardless of their availability. It is therefore important to investigate how students engage with self-paced learning tools and contemplate a holistic framework that can help educators and education researchers devise strategies for supporting and incentivizing students who otherwise may not engage with them effectively as intended.

## II. GOAL OF THIS INVESTIGATION AND RESEARCH QUESTIONS

Our goal was to investigate effectiveness of self-paced tutorials in an algebra-based introductory physics course by evaluating student performance on low-stakes quizzes after students were encouraged to engage with the self-study tools and use the empirical findings as a guide to develop a holistic framework that can help educators and education researchers create effective approaches to incentivize and support productive student engagement with self-study tools. In the empirical study, a large number of students enrolled in the course were encouraged to use interactive tutorials as self-study tools outside of class to prepare for homework and quiz for the coming week. Here we focus on research-validated tutorials involving quantitative problems on three fundamental physics principles that were designed to aid students with diverse prior preparation via a guided approach to learning. Student learning was evaluated by their performance on in-class scaffolded and unscaffolded "pre-quiz" problems (that were identical to the tutorial problems) and "paired" quiz problems (that were given immediately after the pre-quiz problems and were comparable to the tutorial problems in that they involved the same physics principles). The research questions that focus on the empirical study on the impact of these tutorials implemented in that course are:
1. What percentage of students work through optional tutorials when no credit is explicitly associated with their completion?
2. How well do students who worked through the tutorial perform on a "pre-quiz" problem that is identical to the tutorial problem (i.e., can students who worked through the tutorial reproduce the solution of the tutorial problem in a quiz setting)?
3. How does the performance of students who worked through the tutorials compare to those who did not work on them on "pre-quiz" problems that are identical to the tutorial problems (i.e., what is the added benefit of working through a tutorial on a follow up quiz that involves the same problem as the tutorial)?
4. How does the performance of students who worked through the tutorials compare to those who did not work on them on "paired" quiz problems, which are transfer problems on the same underlying concepts given immediately after the "pre-quiz" problems (i.e., what is the added benefit of working through a tutorial on a follow up quiz that involves transfer of learning of a concept from one situation to another)?
5. How does the performance of students who worked through the tutorial as a self-study tool in a large, introductory physics course compare to the performance of the introductory students who engaged with the same tutorial in a one-on-one interview setting on "paired" quiz problems?

The findings from the preceding research questions motivated the development of a holistic framework to help students learn effectively from self-study tools. Below, we first summarize the development, validation and structure of the interactive tutorials, pre-quiz problems, and paired quiz problems. Then we describe the research methodology. Then we discuss findings regarding the effectiveness of these tutorials in one-on-one implementations in individual interviews followed by their implementation as self-study tools as a part of large introductory physics course. We then reflect upon the findings of the empirical investigation and describe a holistic framework that is informed by the findings that can aid in creating learning environments to help all students learn effectively from self-study tools.

## III. DEVELOPMENT, VALIDATION AND STRUCTURE OF THE TUTORIALS

The development of the tutorials was guided by a cognitive apprenticeship learning paradigm [49] which involves three essential components: modeling, coaching, and weaning. In this approach, "modeling" implies that the instructor demonstrates and exemplifies the skills that students should learn (e.g., how to solve physics problems systematically). "Coaching" involves providing students opportunities for practice and guidance so that they are actively engaged in learning the skills necessary for good performance. "Weaning" consists of reducing the support and feedback gradually so as to help students develop self-reliance. The tutorials include modeling via breaking the tutorial problem down into sub-problems dealing with different stages of problem solving and promoting a systematic approach to problem solving. They also include coaching by providing feedback and guidance based on students' difficulties. The coaching and scaffolding are adaptive in that the help provided to students after they answer each multiple-choice sub-problem is tailored to the student's specific difficulty. The adaptive tutorials also involve weaning by gradually providing less scaffolding as student understanding improves and they become more confident in solving the problems on their own. In addition, the tutorials also include reflection problems that require students to apply the concepts learned in different contexts in order to develop self-reliance.

Each tutorial starts with an overarching problem which is quantitative in nature. These problems were chosen to be somewhat more difficult than a typical introductory level physics homework problem on the same physics principle. This increased difficulty was chosen so that the problems could not be solved using a plug and chug approach and would have enough depth to be able to help students learn an expert-like problem solving approach. Figure 1 is an example of one of these overarching problems in the tutorial focusing on Newton's Second Law. Before working through a tutorial, students are asked to attempt the problem to the best of their ability. The tutorial then divides this overarching problem into a series of sub-problems, which take the form of research-guided, conceptual, multiple-choice questions. These sub-problems help students learn effective steps for successfully solving a physics problem, e.g., analyzing the problem conceptually, planning the solution and decision making, implementing the plan, and assessing and reflecting on the problem-solving process. The alternative choices in these multiple-choice questions elicit common difficulties students have with relevant concepts. Incorrect responses direct students to appropriate help in which they are provided suitable feedback and explanations both conceptually and with diagrams and/or appropriate equations to learn relevant physics concepts. The correct responses to the multiple-choice questions advance students to a brief statement affirming their selection followed by the next sub-problem.

Figure 2 shows an example of a sub-problem in the Newton's 2<sup>nd</sup> Law tutorial and adaptive feedback provided to students. The leftmost image in Figure 2 shows the sub-problem in which students are provided an opportunity to determine the components of the gravitational force parallel and perpendicular to the inclined plane. If students select incorrect option D, they are provided with help that focuses on how to determine the components of the gravitational force using a diagram (middle image in Figure 2). If they select option C (which is correct), the feedback confirms that the students' answer choice was correct and gives a reason for why it is correct (rightmost image in Figure 2).

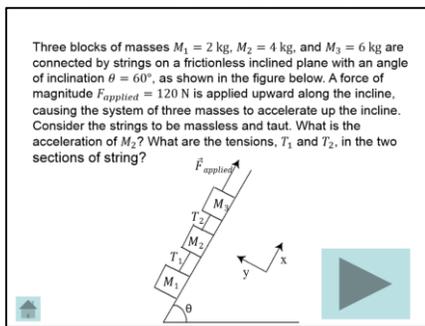

**Figure 1**. The overarching problem in the Newton's 2nd Law tutorial.

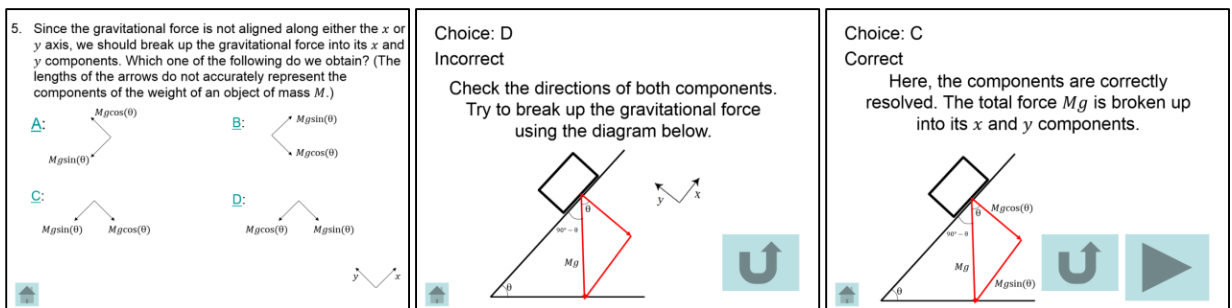

**Figure 2**: Example of a sub-problem from the tutorial focusing on Newton's second law and adaptive support provided if the student provides an incorrect or a correct response. Students can either click on a particular option in the multiple-choice or click on the home button in order to access any of the previous sub-problems and associated help.

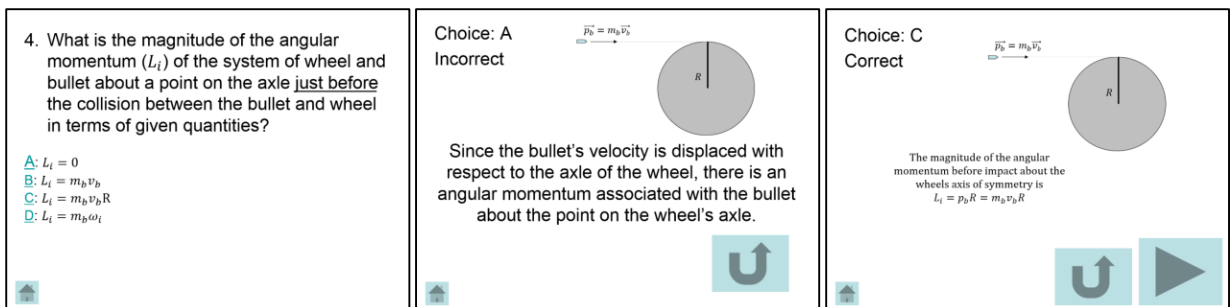

**Figure 3**. Example of a sub-problem from the tutorial focusing on conservation of angular momentum and adaptive support provided if the student provides an incorrect or a correct response.

Figure 3 shows an example of a sub-problem in the conservation of angular momentum tutorial and the adaptive feedback provided to students. The leftmost image in Figure 3 shows the sub-problem in which students are provided an opportunity to determine the magnitude of the initial angular momentum of a particular system. If the students select incorrect answer option A, the tutorial provides feedback that helps students think about the angular momentum associated with the bullet (middle image in Figure 3). If the student selects answer correct option C, the feedback confirms that the student's answer is correct and gives a reason for why it is correct (rightmost image in Figure 3).

After students work on the sub-problems in the tutorials, they answer several reflection sub-problems, which focus on helping students reflect upon what they have learned and apply the concepts learned in different contexts. If students have difficulty answering the reflection sub-problems, the tutorial provides further assistance and feedback in the form of a review of the effective problem-solving approach.

The development of the tutorials went through a cyclic, iterative process. For each tutorial problem, a cognitive task analysis was performed by three graduate student researchers and one professor (all physics education researchers) to break down each tutorial problem into a series of sub-problems dealing with different stages of problem solving that must be answered to solve the tutorial problem. Each sub-problem was then posed as a multiple-choice question. The incorrect options for each multiple-choice question were chosen to emphasize the common difficulties uncovered by having introductory physics students solve similar problems in an open-ended format. Explanations for each multiple-choice option were written and refined, based on one-on-one student interviews, to reinforce student understanding of the reasoning behind the options given and to aid them in repairing and extending their knowledge structure when they select an incorrect option. Using this approach, the initial drafts of the tutorials were created. Each initial draft was revised several times based on interviews with introductory physics students and feedback from graduate students and several professors who were asked to work through them and provide feedback to ensure that they were comfortable with the wording of the sub-problems and progression of the tutorial. During this revision process, the fine-tuned versions of the tutorials were implemented in one-on-one think aloud [50] interviews with introductory physics students and were shown to improve student performance on the paired problems that were developed in parallel with the tutorials.

Twenty such adaptive e-learning tutorials were developed and validated, which cover many topics in introductory physics related to mechanics, electricity, and magnetism [66]. Here we focus on the effectiveness of three of these tutorials intended for use in an introductory mechanics class. The three tutorials selected for this investigation focus on Newton's second law, conservation of energy/work-energy theorem, and conservation of angular momentum.

Each tutorial is also matched with a paired quiz problem that requires the same underlying physics concepts to solve it but is posed in a different context, i.e., transfer of learning [67-87] from the tutorial problem is required in order to solve the paired problem. The paired problems assess whether students have learned to de-contextualize the problem solving approach and concepts learned via the tutorial. The paired quiz problems are open-ended problems that are not broken up into sub-problems. They play an important role in the weaning part of the learning model and assess whether students have developed self-reliance and are able to solve other problems based upon the same underlying concepts as the tutorial without any guidance.

## IV. RESEARCH METHODOLOGY

Below, we describe the methodology for the implementation of the three tutorials in one-on-one implementation with student volunteers and as a self-study tool offered in algebra-based introductory physics courses (taken primarily by undergraduates interested in pursuing bioscience, neuroscience related majors, or medical professions in the future) at the University of Pittsburgh, which is a large, typical state-affiliated university in the US.

### A. One-On-One Implementation

We first determined the effectiveness of tutorials in one-on-one interviews before implementing and evaluating the impact of the tutorials in a large introductory physics course. An email was sent to all students in the course asking for student volunteers to work through physics tutorials in a one-on-one interview setting. The students were paid for their participation and all students that responded to the email were interviewed. Informal questioning of the students regarding their performance in their class exams, homework and quizzes revealed that several of them were performing particularly well and several were struggling in the class. In the courses, typically 10-15% of the students receive a grade of C- or lower, so these recruited students formed a representative group from the class.

In the one-on-one implementation [65], students were observed by a researcher as they worked on the tutorials while thinking aloud and being audio-recorded. The researcher required that the students follow the instructions for working through a tutorial. For example, students had to first attempt to outline the solution to the tutorial problem to the best of their ability before they started the tutorial. Then they were required to answer each sub-problem in the appropriate order. Throughout this one-on-one implementation process, each student was asked to think aloud so that the researcher could understand his/her thought process and the researcher made further record of his observations of each student's interaction with the tool. The researcher remained silent while the students worked unless they became quiet, in which case the researcher prompted students to keep talking. After working through the entire tutorial, the students worked on the corresponding paired problem. This process was repeated with each student for each tutorial.

Twenty-two 2-3 hour long, one-on-one, think-aloud interviews were conducted with volunteers who were either in an algebra or calculus-based introductory physics course. Roughly half of the students were enrolled in an algebra-based course and the other half were enrolled in a calculus-based physics course. These students were paid volunteers who responded to a flyer distributed in the classes and already had traditional classroom instruction related to physics concepts covered in the tutorial. In each interview, between two to three tutorials were covered, depending upon the pace of the student. The interview data was de-identified so it is not possible to match students' interview data with whether they were enrolled in an algebra-based or calculus-based course.

### B. Large scale implementation of the e-learning tutorials as self-study tools

After we found that the tutorials were effective in individual administration, we then implemented them as self-study tools in an algebra-based introductory physics course. The course was a first semester physics course with 385 students (split into two sections). Approximately 60% of the students identified as female, 20% of the students identified as Asian, 65% of the students identified as White, and 13% of the students identified as traditionally disadvantaged and underrepresented minority students The students came from varied backgrounds with approximately 70% of them planning to pursue careers in health professions (e.g., medical or dental careers). Thus, this implementation allowed the researchers to determine the effectiveness of tutorials for students in a course in which researchers had no control over how the tutorials were used by the students as self-study tools. Table I shows a sequence of the self-study tool activities and recitation quizzes in the course.

**Table I.** Sequence of activities involving the self-study tools in an algebra-based introductory physics course.

| In-class | Outside of class | In-class recitation | In-class recitation |
|---|---|---|---|
| Traditional instruction in relevant topics | Worked on tutorial | Scaffolded pre-quiz problem (Multiple-choice tutorial sub-problems) | Paired quiz problem (open-ended transfer problem) |
| | | Unscaffolded pre-quiz problem (Open-ended tutorial problem) | |
| | Did not work on tutorial | Scaffolded pre-quiz problem (Multiple-choice tutorial sub-problems) | |
| | | Unscaffolded pre-quiz problem (Open-ended tutorial problem) | |

Each of the three tutorials were posted on the course website as a self-study tool the week students received classroom instruction in relevant concepts. The tutorials could be used at a student's discretion after the associated physics concepts and principles were introduced in lecture, and the tutorial and the associated homework problems were assigned in the same week. The links to the tutorial were uploaded on the course website but the amount of time each student spent working through them could not be tracked since our learning management system did not have this capability. After students had the opportunity to use each tutorial as a self-study tool, a pre-quiz problem was administered immediately followed by a paired quiz problem in a recitation class in the following week. While the paired quiz problem was the same for all students, those in some recitation classes were randomly administered the scaffolded version of the pre-quiz problem while those in the other recitation classes were administered unscaffolded version of the pre-quiz problem. The scaffolded pre-quizzes consist of multiple-choice questions, structured in the same way as the associated tutorial. In other words, the multiple-choice questions that students answer as part of the scaffolded pre-quiz involve the same questions as the tutorial sub-problems (in the same order as in the tutorial, but students are not provided feedback on whether their choices are correct, unlike the immediate feedback that is available for the sub-problems in the tutorial). Thus, the difference between the tutorials and the scaffolded pre-quizzes is that the tutorial provides adaptive feedback to students after they choose an answer. On the other hand, the scaffolded pre-quiz offers no such feedback or reinforcement when an answer is selected for each multiple-choice question. The unscaffolded pre-quiz is identical to the open-ended tutorial problem and students get no additional scaffolding (problem is not broken into sub-problems). Immediately after students submitted the solution to the pre-quiz problem, they were given the corresponding paired quiz problem; see the Appendix for the paired quiz problems.

The course instructor incentivized the self-study tutorials by telling the students that the tutorials would be helpful for solving assigned homework problems and in-class quiz problems (scaffolded and unscaffolded pre-quiz problems and paired quiz problems) for that week. Although students were made aware that no points would be awarded simply for completing the tutorials, announcements were made in class, posted on the course website, and sent via email informing students that the tutorials were available. All students had sufficient time to complete the quizzes (including both parts: the pre-quiz and the paired quiz problems). Students were given a grade based on their performance on the pre-quiz and paired quiz problems as their weekly quiz grade. On top of each sheet with the paired problem quizzes that were administered in the recitation classes, students were asked the following questions and were assured that their answers to these questions would not influence their score on the quiz:
- Have you worked on the corresponding online tutorial?
- Was the tutorial effective at clarifying any issues you had with the problem covered in the tutorial?
- If the tutorial was ineffective, explain what can be done to make it effective?
- How much time did you spend on the tutorial?

These questions allowed us to separate students into "tutorial" or "non-tutorial" groups, determine the percentage of students who worked through the tutorials, and gain insight into how students who engaged with the tutorials performed on the quizzes (both pre-quizzes and paired quizzes) compared to those who did not. As noted, we could not collect any data via our learning management system to determine whether students opened a tutorial and how long they spent working on it. Therefore, we rely on students' self-reported data regarding whether they engaged with the tutorials and how much time they spent on each.

The purpose of administering the pre-quizzes was two-fold. First, we wanted to examine whether students who worked through a tutorial were able to solve the same tutorial problem successfully in a quiz setting without the adaptive support of the tutorial. The second purpose of the pre-quizzes was to compare the performance of students who worked through the tutorial with those who only worked on the corresponding scaffolded pre-quiz (but did not work through the tutorial) on the paired quiz problems. The pre-quizzes enabled us to evaluate whether students who worked on only the pre-quizzes performed better or worse than those who engaged with a tutorial as a self-study tool on paired problems (which were transfer problems involving the same underlying concepts). In this way, we were able to investigate, in part, whether students who worked through the tutorial were able to transfer their learning to different contexts (compared to those who only worked through a scaffolded pre-quiz that did not include adaptive support).

To compare the performance of students who worked on the tutorials in a one-on-one interview setting with those who used them as a self-study tool, we examined student performance on the paired problems in these two settings. Rubrics were developed by three graduate students and a professor for each paired problem. Once the rubric for grading each paired problem was agreed upon, 10% of the paired problem quizzes were graded independently by three graduate students and a professor with the finalized version of the rubric. When the scores were compared, the inter-rater agreement was better than 90% across all graders. In this way, we were able to examine, in part, the level of student engagement with the tutorials in a one-on-one implementation and a large-scale, self-study implementation.

## V. RESULTS

With regards to Research Question 1, Table II shows that less than 50% of the students self-reported that they worked through each of the tutorials. With regards to Research Question 2, Figure 4 show the average performance of students who worked through the tutorials on the scaffolded and unscaffolded pre-quizzes, respectively. Students who worked through the tutorials did well on the scaffolded pre-quizzes, with an average of above 80% (see the black bars for the scaffolded pre-quizzes in Figure 4). On the other hand, students who worked through the tutorials had an average of less than 70% on the unscaffolded pre-quizzes (see the black bars for the unscaffolded pre-quizzes in Figure 4).

**Table II.** Student responses to the question "Have you worked on the corresponding online tutorial?" in large enrollment classes as a self-study tool.

|  | Yes | No |
|---|---|---|
| Newton's Second Law | 78 | 260 |
| Conservation of Energy | 165 | 217 |
| Conservation of Angular Momentum | 140 | 180 |

With regard to Research question 3, one must compare the black and gray bars for the scaffolded and unscaffolded pre-quizzes in Figure 4. In all cases, the students who worked through the tutorial performed better than those who did not on the scaffolded pre-quizzes (compare the black and gray bars for the scaffolded pre-quizzes) and the unscaffolded pre-quizzes (compare the black and gray bars unscaffolded pre-quizzes). Thus, it appears that there was an added benefit of working through the tutorial on follow up quizzes that involved the same problem as in the tutorial (although the benefit varied for the three problems shown in Figure 4).

We also compared the performance of students who worked through the tutorials to those who did not on "paired" quiz problems, which are transfer problems given immediately after the scaffolded "pre-quiz" problems to determine whether there was any benefit of working through the tutorial on a subsequent quiz that involved transfer of learning (Research question 4). To make this comparison, one must compare the black and gray bars for the paired problems in Figure 4. Students who worked through both a tutorial and an unscaffolded pre-quiz (see the black bars for the paired problems (unscaffolded pre-quizzes) in Figure 4) performed approximately the same as students who only worked on a scaffolded pre-quiz that did not have the adaptive features of the tutorial (see the gray bars for the paired problems (scaffolded pre-quizzes)) on the paired problems that required transfer of learning. In particular, on average, students who worked on the same problem twice (once as a self-study tutorial with adaptive features and once as an unscaffolded pre-quiz) did not perform significantly better on follow-up transfer problems than students who only worked on the problem once as a scaffolded pre-quiz. In addition, the students in the tutorial group who worked on a scaffolded pre-quiz (see the black bars for the paired problems (scaffolded pre-quizzes)) performed approximately the same as the students who only worked on a scaffolded pre-quiz (gray bars for the paired problems (scaffolded pre-quizzes)) on the paired problems. In other words, students who worked on the same problem that provided scaffolding twice (once as a self-study tutorial with adaptive features and once as a scaffolded pre-quiz) did not perform significantly better on follow-up transfer problems than students who only worked on the problem once as a scaffolded pre-quiz. It appears that the students in the tutorial group struggled to transfer their learning to new contexts on the paired problem and did not receive an added benefit from working through either a scaffolded or unscaffolded pre-quiz before solving the paired problems.

We also compared the performance of students in the large scale, self-study implementation of the tutorial with those in the one-on-one implementation of the tutorial in an interview situation on the paired problems to answer Research question 5. Table III shows the average performance of students on the paired problems in the one-on-one interview group. The students in a one-on-one interview setting had an average score above 80% on all the paired problems. However, students in the self-study tutorial group had an average score of approximately 50% on the paired problems (regardless of whether they worked through a scaffolded or unscaffolded pre-quiz). We note that the interview group was comprised of students in algebra-based and calculus-based physics courses, but we were unable to separate out the scores of students in these two groups because the data were de-identified without separating them. However, the standard deviations of the scores of the interview group on the paired problems are small compared to the standard deviations of the scores of the self-study group so a comparison between the interview group and the algebra-based self-study group is appropriate.

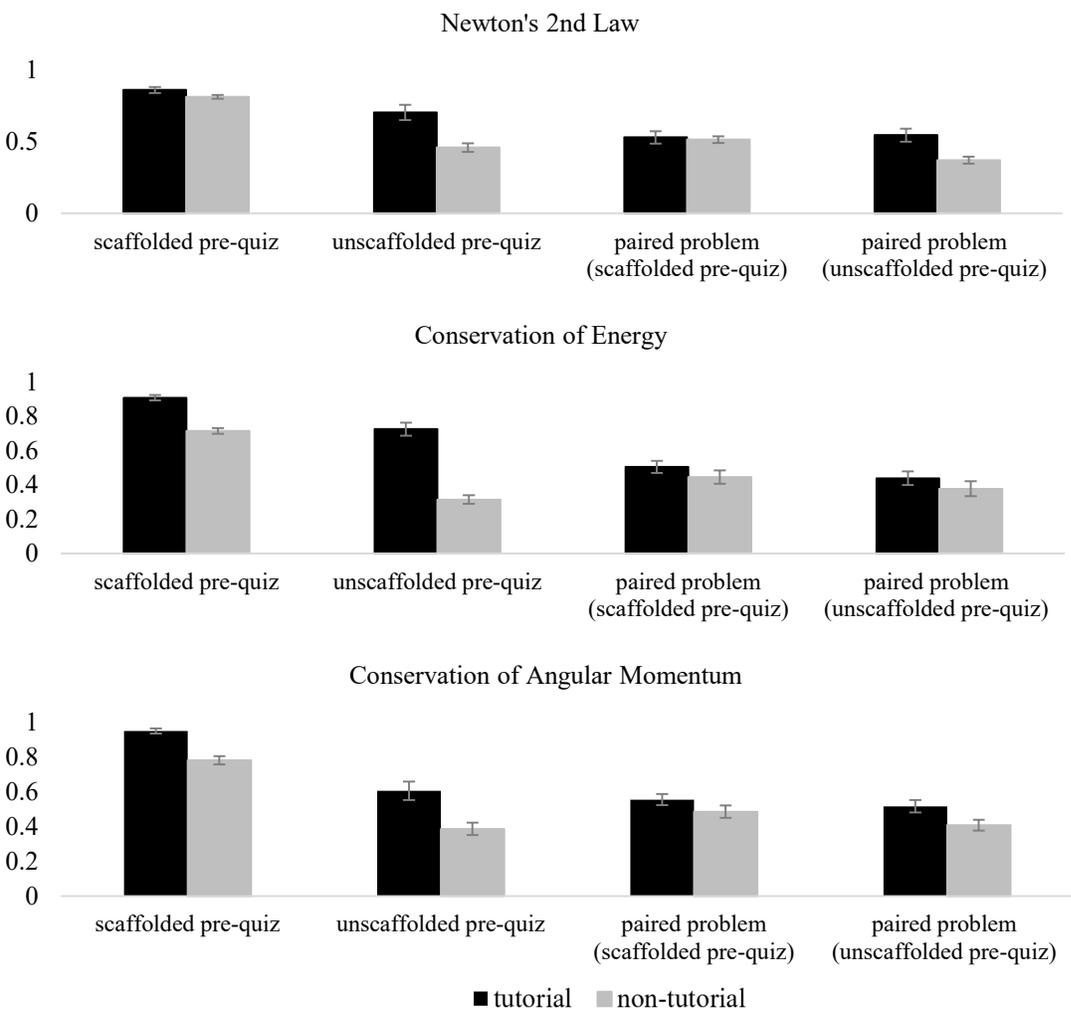

**Figure 4**. Average performance on the pre-quiz problems and paired quiz problems and standard error bars for the students who worked through the tutorials and those who did not.

**Table III.** Average paired problem scores and standard deviations (SD) in one-on-one implementation group.

| Physics Principle | Interview Group (SD) |
|---|---|
| Newton's Second Law | 86.0% (15.9%) |
| Conservation of Energy/work energy theorem | 95.5% (11.8%) |
| Conservation of Angular Momentum | 83.3% (16.0%) |

## VI. DISCUSSION : THEORETICAL FRAMEWORK

In this investigation, we found that less than 50% of the students in the course took advantage of self-study tutorials. This is unfortunate because those who did not work through the tutorials could have benefited from additional help, as evidenced by the low performance of the non-tutorial group on the unscaffolded pre-quizzes and paired problems that required transfer of learning. This finding suggests that many students who can especially benefit from the out of class help via self-paced learning tools in introductory physics courses may not take advantage of them without additional support. What is more, our findings suggest that even those students who reported that they worked through the tutorials may not have engaged with them in an effective manner. For example, while students in the tutorial group performed well on the same tutorial sub-problems in the multiple-choice format (in the scaffolded pre-quiz condition), many of them struggled to solve the same tutorial problem as an open-ended, unscaffolded pre-quiz. In addition, students who worked through the same problem twice (once in the tutorial and once in the scaffolded or unscaffolded pre-quiz) did not perform better than students who worked on the problem once as a scaffolded pre-quiz on paired quiz problems that required transfer of learning. This finding indicates that students who stated that they worked through the tutorial may not have done so effectively and they struggled to transfer their learning to new contexts. Furthermore, we found a dichotomy in the performance of students who worked on the tutorials in a one-on-one setting v. the students who worked on the tutorials in the large-scale implementation as self-study tools. The students in the one-on-one interview setting worked through the tutorial in a deliberate and engaged manner, and they performed well on the paired transfer problems. This finding indicates that the tutorials were effective in helping students learn physics concepts and transfer their learning to new situations when students engaged with them in a deliberate manner. However, students who used the tutorials as self-study tools in a large-scale implementation without supervision did not perform well on the paired transfer problems, indicating that they may not have engaged effectively with the tutorials in a deliberate manner. This dichotomy between the performance of the self-study group and one-on-one implementation group suggests that carefully designed tutorials, when used as intended, can be a powerful learning tool for students across diverse levels of prior preparation and mathematical background, but getting students to engage with them effectively as a self-study tool can be challenging.

The poor performance of the tutorial group on the unscaffolded pre-quizzes and paired problems may be due to students engaging superficially with the tutorials in the large-scale, self-study implementation. Although these students were given instructions on how to work through the tutorial effectively, they could have taken short cuts and skipped sub-problems if they decided not to adopt a deliberate learning approach while using these tools. Indeed, upon examining student comments and other data gathered with their responses to the paired problems in the self-study group, some students explicitly commented that they "skimmed" or "looked over" the tutorials but that type of engagement with the tool may not help them learn deeply and transfer their learning in order to apply the concepts learned to new situations. Additionally, they may not have attempted to first solve the tutorial problem on their own without the scaffolding provided by the tutorial (although explicitly told to do so), even though this step would have allowed them to productively struggle with the problem and prime them to learn from the tutorial [88]. Even some of the students in deliberate one-on-one interviews had to be prompted several times to make a prediction for each sub-problem before selecting an answer rather than randomly guessing an answer without thinking. Furthermore, a detailed look at the performance of students who used the tutorials as self-study tools on the paired problems indeed suggests that many students may have memorized certain equations by browsing over the tutorials, expecting that those equations may help them in solving the in-class quiz problems, instead of engaging with the tools in a systematic manner. Interestingly, in a survey given at the end of the course to students who used them as a self-study tool, a majority noted that they thought that the tutorials were effective even though their performance on the paired problems reflected that they had not learned effectively from them.

The mediocre performance of students in the self-study group on unscaffolded pre-quizzes and paired problems supports the notion that major challenges in implementing tutorials as self-study tools are likely to be students' motivation, self-regulation skills, time-management skills [89-94], and social and environmental factors. It appears that without sufficient support to help students develop self-management and time-management skills and incentives to motivate students to engage with the tutorials, many students may not follow the guidelines for effectively using them. The haphazard use of these tools can reduce their effectiveness significantly. Many students have difficulty realizing that much of the value to be gained from these tools depends on interacting with them in a prescribed manner. Therefore, it is important for educators and education researchers to contemplate how to provide appropriate support in order for students to benefit from learning tools—especially if they are research-validated, adaptive tools that have been found effective in one-on-one implementation. Below, we propose a theoretical framework for this purpose.

The framework, called Self-study for Engaged Learning Framework (SELF) (see Figure 5), is a holistic framework which suggests that instructional design and learning tools, their implementation, student characteristics, and social and environmental factors collectively determine how effectively students engage with and learn from instruction in a particular course [95,96]. The framework consists of four quadrants, and all of them must be considered holistically to help a diverse group of students learn effectively. The horizontal dimension involves the characteristics of learning tools (e.g., how the tool provides efficiency and innovation in learning and builds on students' prior knowledge) and students (e.g., students' prior preparation and motivational characteristics), both of which should be taken into account when developing effective learning tools. The vertical dimension involves internal and external characteristics of the learning tools and the students. This dimension focuses on how the characteristics of the learning tools and students as well as the environments in which the tools are implemented are important to consider when helping students engage with and learn from these learning tools. The internal characteristics of the tool pertain to the tool itself (e.g., whether it includes formative assessment). The external characteristics of the tool pertain to how the tool is implemented (e.g., whether the tool is framed appropriately to get student buy in). The internal characteristics of the students pertain to, e.g., their prior preparation, motivation, goals, and epistemological beliefs. The external characteristics of the students pertain to social and environmental factors such as support from mentors and advisors and time management in order to balance multiple demands of everyday life.

Most of the self-study tools so far have mainly taken into account the upper two quadrants of the framework in Fig. 5. The upper-left quadrant involves the learning tool characteristics that directly focus on knowledge and skills to be learned. For example, the cognitive apprenticeship model [64] can be used to develop adaptive tools to promote mastery of the material for a variety of students. These materials, when developed carefully via research in education, can provide scaffolding support to a variety of students. In order to make the self-study tools effective, educators often consider the user characteristics in the upper-right quadrant [53-63]. The various models of learning lead to similar conclusions about how to connect user characteristics with the characteristics of the learning tools (i.e., how to connect Factors I and II). For example, Schwartz, Bransford, and Sears' preparation for future learning model [97] emphasizes that in order for students to engage appropriately with learning tools, there should be elements of both efficiency and innovation embedded in the instructional tools and design. One interpretation of this model is that if the students are asked to engage with learning tools that are too efficient, students will disengage. On the other hand, if the learning tools are too innovative, students will struggle so much while engaging with them that they will get frustrated and give up. Thus, the learning tools should have appropriate blending of both efficiency and innovation so that students engage and struggle productively while learning [98,99]. Ensuring the appropriate balance of efficiency and innovation requires that the learning tool takes into account students' prior knowledge (Factor II). In addition, effective learning tools have formative assessment built into them so that students can receive feedback and evaluate their own learning as they make progress. If the learning tool is administered online, the developers of the tool can take into account research on how students interact with online tools [100-102]. In particular, these tools should take into account, e.g., question timing and pathways. Furthermore, research has shown that the ways in which online mastery homework is implemented can improve student engagement [103] and providing scaffolding supports can help students achieve mastery of the material more quickly [104]. Since student characteristics within a particular course vary, carefully designed adaptive interactive tools can provide appropriate balance of innovation and efficiency for a variety of students [105-108]. Students who are lacking some elements of prior knowledge can benefit from a carefully designed learning tool which involves formative assessment, allows students to make mistakes but learn from them and try again, and scaffolds their learning by providing elements of both efficiency and innovation [109-115].

In the study described here, the tutorials included considerations of Factors I and II. For example, the tutorials were inspired by the cognitive apprenticeship learning model. They also provided an opportunity for productive struggle in that they specifically encouraged students to work on each tutorial problem before starting to work on each of the sub-problems. The act of struggling with the tutorial problem can help students connect what they are learning with their prior knowledge and aid in learning. Additionally, struggling with the tutorial problem before engaging with the tutorial may increase students' motivation to engage deliberately with the tool as prescribed. However, the tutorials could also be improved based upon consideration of Factors I and II. For example, the longer tutorials were more complex since they either involved application of more than one physics principle or application of the same principle (Newton's second law) in different contexts. These longer tutorials are useful for helping students develop both content knowledge and skills to solve complex problems. Based upon considerations of Factors I and II, one strategy that may make them more effective is to break the multi-principle tutorials into single principle tutorials and then have another multi-principle tutorial that combines the learning in those single-concept tutorials [115]. Since students would have been exposed to the individual concepts in various single-concept tutorials, they will be more likely to effectively engage with the multi-principle tutorial that consolidates those principles into a more complex problem.

The study suggests that carefully developed adaptive interactive learning tools, which take into account students' prior knowledge, will not necessarily help them learn if students are not adequately supported and incentivized and do not take advantage of the self-study tools to learn in an effective manner. While the top two quadrants in our framework are often considered in the development of learning tools (although even there, motivational characteristics of students are not adequately accounted for), the lower two quadrants of the SELF have mostly been ignored. But as the study presented here indicates, these lower two quadrants are likely to play a critical role in whether students, especially those with inadequate prior knowledge in need of learning via self-study tools, take advantage of them. The lower right quadrant or Factor IV focuses on external student-environment interaction characteristics, e.g., how students interact with their surroundings, how they manage their time, and how they regulate themselves. Factor IV also involves support that students may receive from their environments such as help from family, advisors, mentors and counsellors to manage their time better and engage in learning effectively. In particular, in our study, students' engagement with the learning tools may have been impacted by whether they have self-regulation and time-management skills, family encouragement, and support from advisors and counselors.

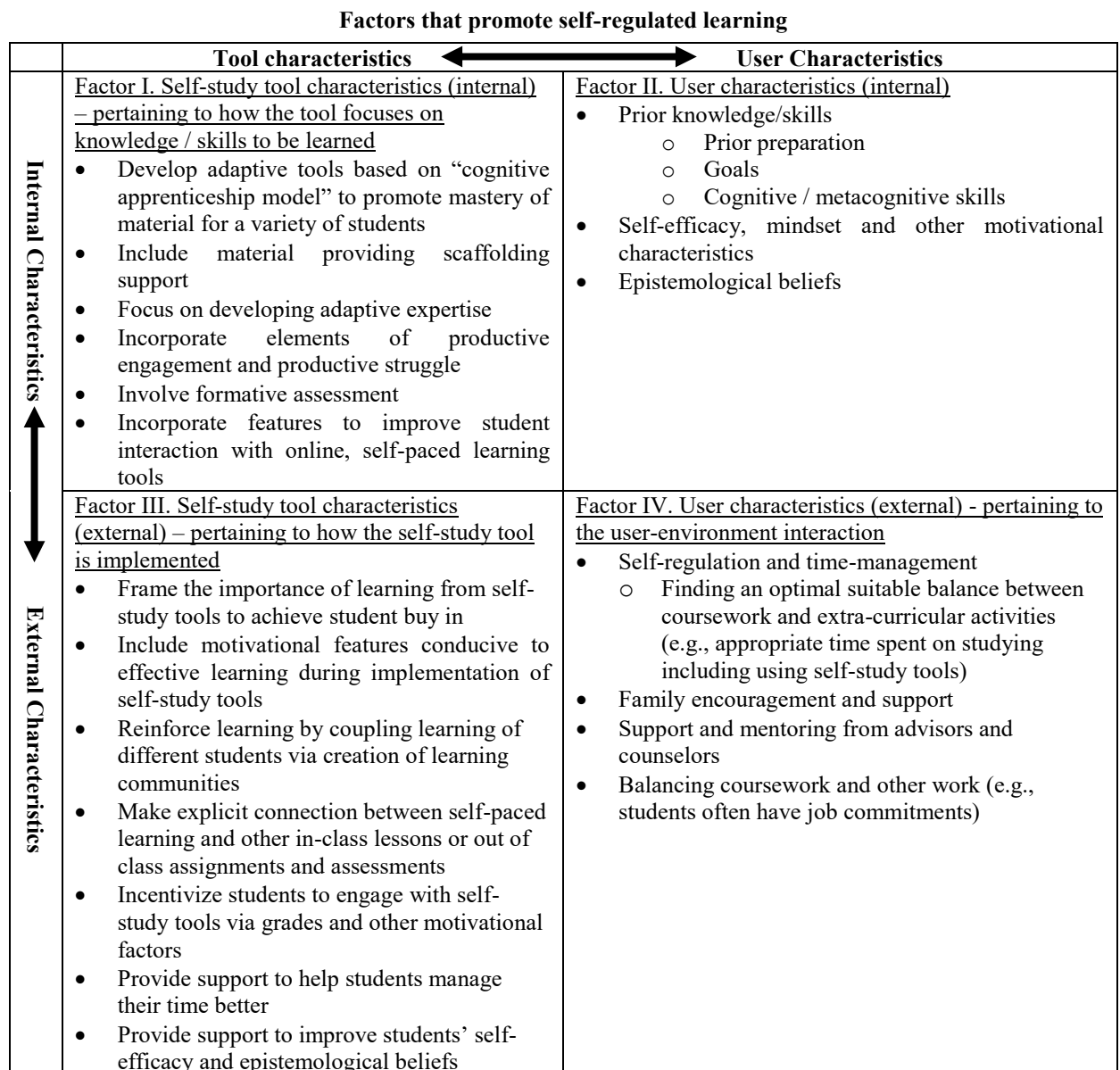

**Figure 5**. Self-study for Engaged Learning Framework (SELF)

**Factors that promote self-regulated learning**

|  | Tool characteristics ⟷ | User Characteristics |
|---|---|---|
| **Internal Characteristics** | Factor I. Self-study tool characteristics (internal) – pertaining to how the tool focuses on knowledge / skills to be learned<br>• Develop adaptive tools based on "cognitive apprenticeship model" to promote mastery of material for a variety of students<br>• Include material providing scaffolding support<br>• Focus on developing adaptive expertise<br>• Incorporate elements of productive engagement and productive struggle<br>• Involve formative assessment<br>• Incorporate features to improve student interaction with online, self-paced learning tools | Factor II. User characteristics (internal)<br>• Prior knowledge/skills<br>   ◦ Prior preparation<br>   ◦ Goals<br>   ◦ Cognitive / metacognitive skills<br>• Self-efficacy, mindset and other motivational characteristics<br>• Epistemological beliefs |
| **External Characteristics** | Factor III. Self-study tool characteristics (external) – pertaining to how the self-study tool is implemented<br>• Frame the importance of learning from self-study tools to achieve student buy in<br>• Include motivational features conducive to effective learning during implementation of self-study tools<br>• Reinforce learning by coupling learning of different students via creation of learning communities<br>• Make explicit connection between self-paced learning and other in-class lessons or out of class assignments and assessments<br>• Incentivize students to engage with self-study tools via grades and other motivational factors<br>• Provide support to help students manage their time better<br>• Provide support to improve students' self-efficacy and epistemological beliefs | Factor IV. User characteristics (external) - pertaining to the user-environment interaction<br>• Self-regulation and time-management<br>   ◦ Finding an optimal suitable balance between coursework and extra-curricular activities (e.g., appropriate time spent on studying including using self-study tools)<br>• Family encouragement and support<br>• Support and mentoring from advisors and counselors<br>• Balancing coursework and other work (e.g., students often have job commitments) |

The question then boils down to whether there are external characteristics (that pertain to how the tool is implemented) that can assist students who otherwise may not engage with them effectively due to personal constraints. This external additional support from educators for self-regulation and effective use of the learning tool is included in the lower left quadrant (Factor III) and focuses on providing motivation for engagement and support, taking into account the user characteristics and user-environment interactions. Consideration of the various types of support in Factor III during the implementation of the learning tools is critical to ensure that most students engage with them effectively. In our study, students may have engaged more effectively with the tutorials if elements from Factor III were included in the implementation of the tools. Indeed, since many students may have disengaged with the longer tutorials while using them as a self-study tool, finding better ways to keep students motivated while working through them should be a high priority, rather than only developing shorter tutorials focused on one physics concept/principle [112]. Furthermore, learning tool developers or implementers can consider embedding modules that focus on motivating students to engage with them effectively and strive to get buy-in from students by having them think carefully about why they should engage effectively with them and how they can help them in the long term. Similarly, students who are struggling to manage their time well can be provided some modules to guide them in making a better daily schedule which includes time to learn from those tools (e.g., once students have made a schedule that includes time slots for learning from the tools, electronic notifications can remind them of their schedule as needed). In addition, making explicit connections between self-paced learning and other in-class lessons or out of class assignments and assessments can also help students engage with them more effectively.

Moreover, students who have difficulty engaging with the self-study tools due to lack of self-efficacy or unproductive epistemological beliefs [117-119] about learning should be guided to help them develop self-efficacy [53,54] and productive epistemological beliefs. For example, a short online intervention has been shown to improve student self-efficacy significantly [55]. Similarly, students who have unproductive epistemological beliefs (such as physics is just a collection of facts and formulas, only a few smart people can do physics, and they should just memorize physics formulas and regurgitate them) are unlikely to productively engage with the self-study tools designed to help them develop expertise in physics. It is important to address these issues in order to ensure that students who are most in need of learning using these tools benefit from them and retain what they learn [120-129]

Another factor (see lower left quadrant of the framework) that may help students engage with these tools effectively is creating learning communities of students who are all expected to learn from these tools and then have them engage in some follow up activities in a group environment (this group work can be done electronically or physically depending on the constraints of the class). In this way, individual students may feel more accountable to their group members and effectively use self-study activities to prepare for the group activities. For example, in the study discussed here, encouraging and incentivizing students to work in these types of learning communities could have aided students in engaging with the tutorials more effectively. In particular, if students knew that they were assigned to work with a group on a complex physics problem, they may have had more motivation to work through the tutorials individually in order to prepare for the group work.

Moreover, research has shown that grade incentives can increase student engagement and improve learning [131,132] (see lower left quadrant of the framework). For example, to help students engage effectively with the tutorials, an instructor could incentivize participation in learning via better grade incentives to ensure that students work on them as prescribed. Also, if students work systematically on them and are engaged throughout, they are unlikely to have cognitive overload [133,134] since learning is scaffolded throughout and one sub-problem builds on another. One motivating factor would be to award course credit to students based on their answers to each sub-problem with decreasing score if they guess multiple times. This strategy might be more successful at motivating them to answer each sub-problem carefully (as opposed to randomly guessing an answer) while working through the tutorial. In addition, it is possible that asking students in the study described here to submit a copy of each correct answer to the sub-problem each week and explain why each alternative choice to each sub-problem is incorrect as part of their homework may increase their motivation to engage with these self-study tools (especially because students have many conflicting priorities for their time and they may not engage with them effectively if working through them is not *directly* tied to the grade).

We note that in our framework, Factor III may also impact Factor IV. When students are motivated to think about the importance of using the learning tools, are given credit to work through them, work in learning communities that keep each student accountable while providing mutual support, and can discern the connections between the learning tools and in-class assignments, homework, and quizzes, they may manage their time more effectively. Connecting the learning tool content to real-world applications can also increase student motivation to learn from them. It is also important to note that Factors I and III can impact Factors II and IV so we cannot disentangle any of these factors. Students who lack some prior preparation may also fall behind and have difficulty in managing their time effectively.

Many students who can greatly benefit from using the tools may lack time-management skills. Other students may not have good prior preparation but they may have good time-management skills. In all these cases, in order to help students learn effectively from these tools, Factor III affordances should outweigh the constraints. Therefore, consideration of Factor III, which is often not deliberated on very carefully by educators, is critical.

## VII. SUMMARY AND CONCLUSIONS

The empirical study investigated the impact of tutorials implemented as self-study tools in a large, introductory physics course via student performance on subsequent pre-quiz problems and paired quiz problems administered in recitation classes. The lack of effectiveness when students used the tutorials as a self-study tool in terms of reproducing the solutions to the tutorial problems (pre-quizzes) and transferring learning to new contexts (paired quizzes) is likely due to many students engaging with the tutorial in a superficial manner. Despite the encouragement from the instructor, it was difficult to ensure that students used an effective approach while working through the self-paced tutorials on their own time. This conclusion is supported by the notably higher scores on the paired quizzes in the one on one interview group in which the interviewer could ensure that the correct protocol was used by all students.

Our empirical investigation suggests that despite the ease with which students can access the e-learning tutorials, there are challenges in ensuring that students, especially those who need out-of-class scaffolding support, use them effectively as a self-study tool as intended. In particular, students who do not use the deliberate approach outlined for them when engaging with self-study tools are unlikely to benefit significantly from them. The implications of these findings may extend to other self-study tools.

We used the empirical findings as a guide to develop the SELF framework that emphasizes that in order for students with diverse prior preparations to benefit from self-study tools, educators must holistically consider various internal and external affordances/constraints and include various facets of student engagement with self-study tool in the development and implementation of those tools. In particular, students interacting with even the best designed self-study tools are likely to do so in ways other than those prescribed explicitly to maximize their benefit, which can greatly diminish the tools' effectiveness. This limitation is inherent to self-study tools that have no means of regulating the ways in which the student interacts with them unless, e.g., issues discussed in the SELF framework in the lower left quadrant are explicitly incorporated. The SELF framework also emphasizes that a lack of sufficient motivation, self-regulation, and time-management skills while engaging in learning using self-study tools may turn out to be a major impediment in students benefiting from them unless explicit strategies are employed to ensure engaged learning.

## ACKNOWLEDGEMENTS

We thank the National Science Foundation for award PHY-1806691 and the University of Pittsburgh physics education research group for their help with the rubric development and grading to establish inter-rater reliability.

# APPENDIX

**Paired problem for Newton's Second Law Tutorial**

A basket is on a frictionless inclined plane and is connected to a cylinder by a massless rope and massless frictionless pulley as shown below. If the basket has a mass of 2 kg and can hold 35 apples weighing 0.2 kg each when full, what must be the mass of the cylinder such that a basket full of 35 apples and the cylinder can remain at rest when they are arranged as shown below and are initially at rest?

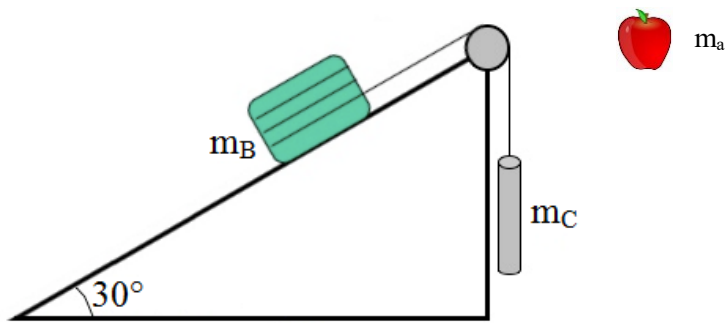

**Figure 6:** Figure for paired problem for "Newton's Second Law" tutorial

**Paired problem for Conservation of Energy Tutorial**

In Figure 7, a horizontal spring with spring constant $k_1 = 800$ N/m is compressed 20 cm from its equilibrium position by a 4 kg block. Then the block is released. What is the maximum distance L that the block travels up a 30° inclined plane as shown in Figure 7? Assume that the track is frictionless.

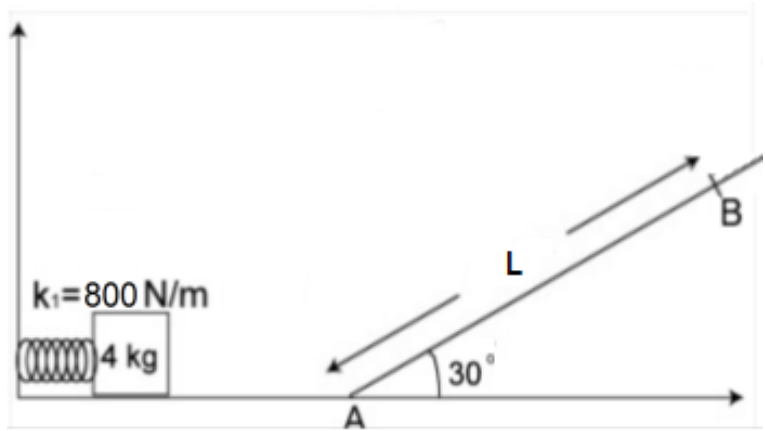

**Figure 7:** Figure for paired problem for "Conservation of Energy" tutorial

**Paired problem for Conservation of Angular Momentum Tutorial**

Suppose that a merry-go-round, which can be approximated as a disk, has no one on it, but it is rotating about a central vertical axis at 0.2 revolutions per second. If a 100kg man quickly sits down on the edge of it, what will be its new speed? (A disk of mass m and radius R has a moment of inertia I=(1/2)mR$^2$ , mass of merry-go-round = 200kg , radius of merry-go-round=6m)